\documentclass[]{elsart}

\usepackage{graphicx}
\usepackage{dcolumn}
\usepackage{amsmath}
\usepackage[]{fontenc}

\begin{document}
\begin{frontmatter}
\title{A new measurement of the $\Xi _{c}^{0}$ lifetime}

The FOCUS Collaboration\footnote{see http://www-focus.fnal.gov/authors.html for
additional author information.}
\author[ucd]{J.~M.~Link}
\author[ucd]{M.~Reyes}
\author[ucd]{P.~M.~Yager}
\author[cbpf]{J.~C.~Anjos}
\author[cbpf]{I.~Bediaga}
\author[cbpf]{C.~G\"obel}
\author[cbpf]{J.~Magnin}
\author[cbpf]{A.~Massafferri}
\author[cbpf]{J.~M.~de~Miranda}
\author[cbpf]{I.~M.~Pepe}
% ...

\address[ucd]{University of California, Davis, CA 95616}
\address[cbpf]{Centro Brasileiro de Pesquisas F\'isicas, Rio de Janeiro, RJ, Brasil}
\address[cinv]{CINVESTAV, 07000 M\'exico City, DF, Mexico}
\address[cu]{University of Colorado, Boulder, CO 80309}
\address[fnal]{Fermi National Accelerator Laboratory, Batavia, IL 60510}\nopagebreak
\address[fras]{Laboratori Nazionali di Frascati dell'INFN, Frascati, Italy I-00044}
\address[ui]{University of Illinois, Urbana-Champaign, IL 61801}
\address[iu]{Indiana University, Bloomington, IN 47405}
\address[korea]{Korea University, Seoul, Korea 136-701}
\address[kp]{Kyungpook National University, Taegu, Korea 702-701}
\address[milan]{INFN and University of Milano, Milano, Italy}
\address[nc]{University of North Carolina, Asheville, NC 28804}
\address[pavia]{Dipartimento di Fisica Nucleare e Teorica and INFN, Pavia, Italy}
\address[pr]{University of Puerto Rico, Mayaguez, PR 00681}
\address[sc]{University of South Carolina, Columbia, SC 29208}
\address[ut]{University of Tennessee, Knoxville, TN 37996}
\address[vu]{Vanderbilt University, Nashville, TN 37235}
\address[wisc]{University of Wisconsin, Madison, WI 53706}

\date{\today}

\begin{abstract}
{\normalsize Using data collected by the Fermilab experiment FOCUS, we
measure the lifetime of the charmed baryon \( \Xi _{c}^{o} \) using the decay 
channels \( \Xi _{c}^{o}\rightarrow \Xi ^{-}\pi ^{+} \)
and \( \Xi _{c}^{o}\rightarrow \Omega ^{-}K^{+} \). From a combined sample
of \( 110\pm 17 \) events we find 
\( \tau (\Xi _{c}^{o})~=~118^{\,+\,14}_{\,-\,12}~\pm5~\)~fs, 
where the first and second errors are statistical and systematic, respectively.}{\normalsize \par}
\end{abstract}

\end{frontmatter}

\section{{\normalsize Introduction}\normalsize }

{\normalsize While the lifetime hierarchy of the weakly decaying charm mesons is well
established experimentally \cite{mesont} and in fairly good agreement with
theoretical calculations \cite{baryonst2}, the pattern of lifetimes in the charm baryon
sector still needs to be established both experimentally and
theoretically \cite{baryonst1}.
While the \( \Lambda _{c}^{+} \)
lifetime is known to an accuracy of about 2\% \cite{lamb3,lamb1,lamb2}, and
recently the \( \Xi _{c}^{+} \) lifetime has been determined to an accuracy
of 5\% \cite{eddy}, the \( \Xi _{c}^{0} \) and \( \Omega _{c}^{0} \)
lifetimes remain known to only 20\% and 30\%, respectively.}{\normalsize \par}

{\normalsize The lifetime of the \( \Xi _{c}^{0} \) has been measured by two experiments.
CERN experiment NA32 found a lifetime of \( 82^{+59}_{-30} \)~fs
using four events reconstructed in the decay channel \( \Xi _{c}^{0}\rightarrow pK^{-}\bar{K}^{*0} \) \cite{accmor}.
Fermilab experiment E687 \cite{e687} measured a \( \Xi _{c}^{0} \)
lifetime \cite{tau_E687} of \( 101^{+25}_{-17} \) (stat) \( \pm 5 \) \rm (sys)~fs
using 42 events from the decay \( \Xi _{c}^{0}\rightarrow \Xi ^{-}\pi ^{+} \).}{\normalsize \par}

{\normalsize FOCUS (Fermilab E831) is a general purpose experiment investigating charm physics.
The charmed particles are produced by the interaction of a photon beam on a segmented BeO target. The
average energy of the beam for the data collected for this measurement is about 180~GeV.
The track reconstruction
is accomplished by two silicon vertex detectors \cite{nimssd,will} (TS and SSD) in
the target region and by five multi-wire proportional chambers (MWPC) downstream
of the target region. The charged momenta are determined by a measurement
of the bending angles in two magnets of opposite polarity. Due to the excellent
separation between production and decay vertices provided by the silicon detectors,
we achieve an average proper time resolution of about 50~fs for the decay channels
used in this analysis. Charged particle identification of hadrons is performed with three multi-cell threshold
\v{C}erenkov counters \cite{jim}. }{\normalsize \par}

\section{{\normalsize Reconstruction of hyperons \protect\( \Xi ^{-}\protect \) and
\protect\( \Omega ^{-}\protect \)}\normalsize }

{\normalsize We reconstruct the decay modes \( \Xi _{c}^{0}\rightarrow \Xi ^{-}\pi ^{+} \)
and \( \Xi _{c}^{o}\rightarrow \Omega ^{-}K^{+} \).
A detailed description of the reconstruction method 
and the mass spectra of the hyperons \protect\( \Xi ^{-}\protect \) and \protect\( \Omega ^{-}\protect \)
is in Reference \cite{john}.
We reconstruct the \( \Xi ^{-} \) decay channel \( \Lambda ^{0}\pi ^{-} \)
and the \( \Omega ^{-} \) decay channel \( \Lambda ^{0}K^{-} \). While these
decay topologies are very similar, we can easily distinguish between them using
\v{C}erenkov identification of the final state particles and by requirements
of the reconstructed invariant masses.
We reconstruct the entire decay chain of these two modes.
The \( \Lambda ^{0} \) is identified by its decay into a proton and an oppositely
charged pion. The tracks of these charged daughters are used to form the decay
vertex of the \( \Lambda ^{0} \) and to determine its flight direction and
momentum. This direction is used together with the momentum of either the
\( \pi ^{-} \) or the \( K^{-} \), to determine the decay vertex and momentum
of the hyperon. The hyperon vertex must lie upstream of the \( \Lambda ^{0} \)
vertex. Finally, the position and momentum vector of the hyperon is matched
to a SSD track.}{\normalsize \par}

{\normalsize For both modes we compute the invariant mass of the two body combination
and select the events which are in the correct hyperon mass region. For the \( \Xi ^{-} \)
we require $1.312~\textrm{GeV}/c^2<M({\Lambda ^{0}\pi ^{-}})<1.330~\textrm{GeV}/c^2$
and for the \( \Omega ^{-} \) we require 
\( 1.665~\textrm{GeV}/c^2<M({\Lambda ^{0}K^{-}})<1.680~\textrm{GeV}/c^2\).}{\normalsize \par}
\section{{\normalsize Reconstruction of \protect\( \Xi _{c}^{0}\protect \) candidates}\normalsize }
{\normalsize The \( \Xi _{c}^{0} \) candidates are reconstructed using a candidate
driven vertexing algorithm \cite{e687}. Briefly, combinations of tracks are used to form
a secondary vertex which in turn must point within errors to a primary production
vertex. Pairs of \( \Xi ^{-} \) (\( \Omega ^{-} \)) candidates and oppositely
charged pions (kaons) are intersected to form a secondary vertex and a \( \Xi _{c}^{0} \)
mass. Several criteria are used to reject the background contributions. Any track consistent with an \( e^{+}e^{-} \)
pair hypothesis is rejected. A confidence level (CLS) 
for the secondary vertex hypothesis from the intersection of the two tracks
is calculated, and a minimum cut is applied, the magnitude of which
depends on the decay mode.
The \( \Xi _{c}^{0} \) decay products are used to construct the \( \Xi _{c}^{0} \) momentum vector, which
is combined with at least two additional silicon tracks to form the primary vertex, which must lie within the
target region. A confidence level (CLP)
for the primary vertex is calculated, and the combination is rejected
if CLP~$<$~1\%. }{\normalsize \par}

{\normalsize We require a minimum value for the significance of the separation between
the secondary and primary vertex, given by \(  \ell/\sigma _{\ell} \), where \( \ell \) is
the distance between the two vertices and \( \sigma _{\ell} \) is the uncertainty
on \( \ell \). The \v{C}erenkov particle identification uses requirements
on the variables W$_e$, W$_{\pi}$, W$_K$, and W$_p$,}{\normalsize  which represent the negative
log likelihood for the hypothesis that a particular track is an electron, a
pion, a kaon, or a proton, respectively. The difference between two of these variables
represents the ratio between the two probabilities. The pion consistency of
a track is defined by a requirement on the variable \( picon=W_{\rm min}-W_{\pi } \),
where \( W_{\rm min} \) is the minimum likelihood of the four corresponding hypotheses.
The specific requirements for the two decay modes are now discussed.}{\normalsize \par}

{\normalsize For the \( \Xi^- \pi^+  \) mode, we apply the detachment cut \( \ell/\sigma _{\ell}>3 \)
and we require CLS~$>$~3\%. During the data collection the tracking system
was improved with the introduction of a silicon vertex detector in the target
region; we require \( \sigma _{t}<150 \)~fs and \( \sigma _{t}<80 \)~fs where
\( \sigma _{t} \) is the error on the decay time of the \( \Xi _{c}^{0} \),
respectively, for data collected before and after the introduction of this detector.
The pion (decay product of the \( \Xi _{c}^{0} \)) is required to have} \emph{\normalsize picon}{\normalsize ~\( >-10 \)
and a momentum greater than 7~GeV/c. }{\normalsize \par}

{\normalsize For the \( \Omega^- K^+ \) mode, the detachment cut is \(  \ell/\sigma _{\ell}>0.5 \).
For the secondary vertex we require CLS~$>$~1\%. To identify kaons, we require \(
W_{p}-W_{K}>-0.1 \) and \( W_{\pi }-W_{K}>1 \),
comparing the kaon hypothesis to the proton and the pion hypothesis. Since \( \Xi ^{-} \) particles
are produced more copiously than \( \Omega ^{-} \) particles, we compute the
invariant mass for the \( \Omega ^{-} \) when it is calculated as a \( \Lambda ^{0}\pi ^{-} \)
combination. To eliminate \( \Xi ^{-} \) contamination under the \( \Omega ^{-} \)
mass, we require \( M(\Lambda ^{0}\pi ^{-})>1.375~\textrm{GeV}/c^2 \).  }{\normalsize \par}

\begin{figure}[p!hb]
{\par\centering \resizebox*{12cm}{10cm}{\includegraphics{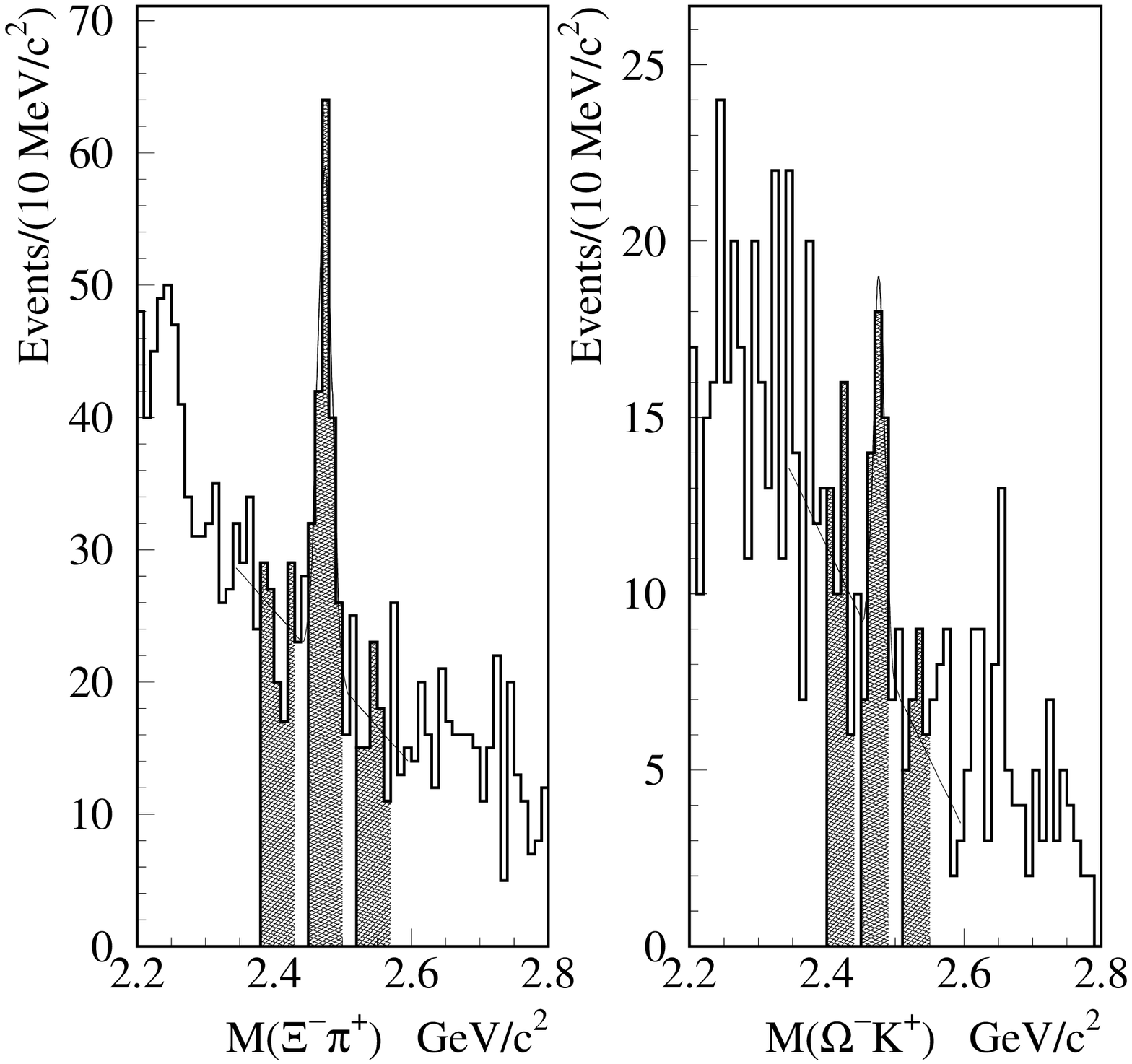}} \par}

\caption{\label{masses}Invariant mass distributions for each mode. The fit 
uses a Gaussian for the signal and a straight line for the background.
The signal region is defined by the central shaded region. The sideband region, defined by the two symmetrical
shaded regions, contains the events used to reproduce the decay time of the background events in the signal
region.}
\end{figure}

{\normalsize Figure \ref{masses} shows the invariant mass plots for the two modes. The fit is to a
Gaussian for the signal and a straight line for the background. We do not use the fit information
to perform the lifetime measurement.}{\normalsize \par}

\section{{\normalsize The lifetime measurement}\normalsize }

\begin{figure}
{\par\centering \resizebox*{10cm}{10cm}{\includegraphics{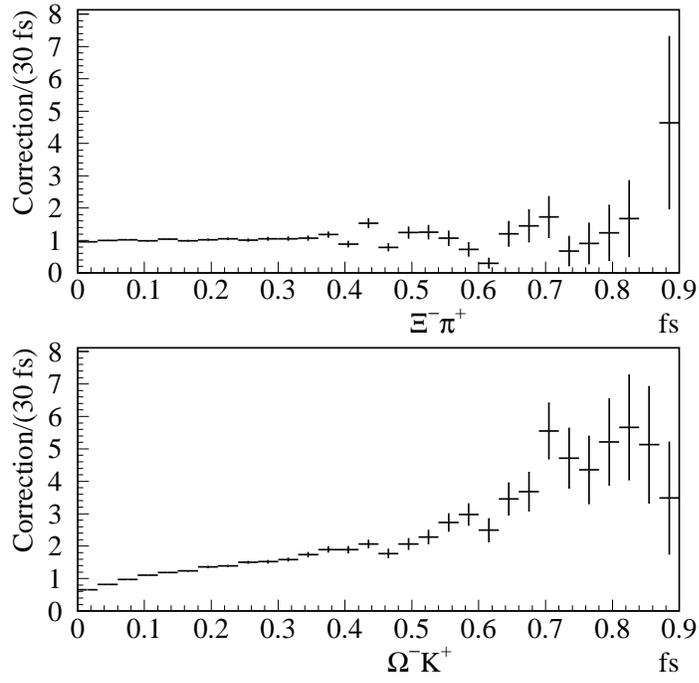}} \par}

\caption{\label{fts}Monte Carlo correction function $f(t')$ for each decay mode.}
\end{figure}

\begin{figure}
{\par\centering \resizebox*{10cm}{10cm}{\includegraphics{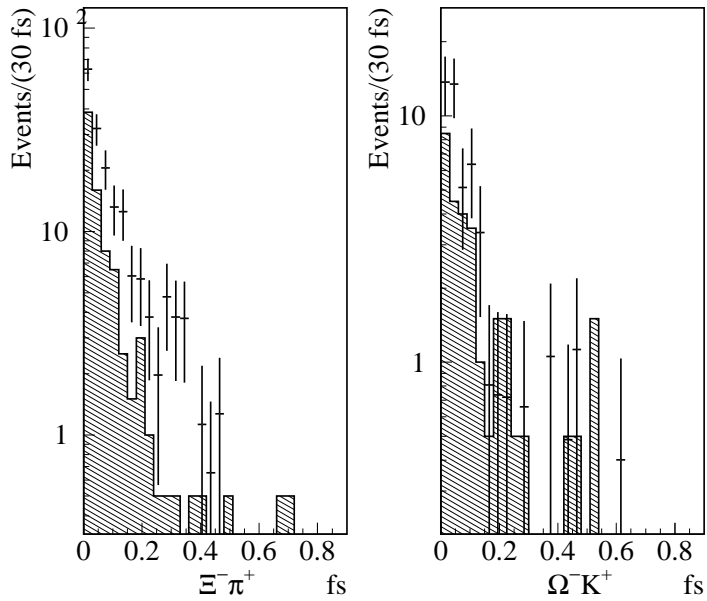}} \par}

\caption{\label{times}Reduced proper time distributions, corrected by the Monte Carlo correction function,
 for each decay mode in the signal region and the
sideband region (shaded).}
\end{figure}

{\normalsize We use a binned maximum likelihood fit method to measure the lifetime.
The fitted histogram is the reduced proper time distribution: \( t'=(\ell-N\sigma _{\ell})/(\beta \gamma c) \),
where $N$ is the vertex detachment cut, and \( \beta \gamma =p(\Xi _{c}^{0})/M(\Xi _{c}^{0}) \).
The fit is performed on the data in the signal region}  {\normalsize
of the invariant mass distribution: the mass window within \( \pm 2~\sigma  \)
of the \( \Xi _{c}^{0} \) nominal mass, where $\sigma$ is the width of the Gaussian fit of the
invariant mass distribution. The \( t' \) distribution of the
signal region differs from the expected behavior \( e^{-t'/\tau } \) due to the presence
of background events. We assume that the background events in the signal region
have the same time distribution as the events of the sidebands }{\normalsize 
(located between 5 and 9~\( \sigma  \) away from the \( \Xi _{c}^{0} \)
mass). If $S$ is the total number of signal events in the signal region, and $B$ is the total
number of background events in the signal region, the expected number of events \( n_{i} \)
in the \( i^{\rm th} \) bin of the \( t' \) distribution is: 
\[
n_{i}=S\frac{f(t_{i}')e^{-t_{i}'/\tau }}{\sum _{i}f(t_{i}')e^{-t_{i}'/\tau }}+B\frac{b_{i}}{\sum _{i}b_{i}}\]
where \( b_{i} \) is the number of events in the \( i^{\rm th} \) bin of the
sideband \( t' \) distribution. The Monte Carlo correction
function $f(t')$ takes into account detector acceptance, the efficiency of the
selection cuts, and absorption of the daughter particles as a function of the
reduced proper time. 
Figure \ref{fts} shows the correction functions for the
two modes; due to the lower detachment cut we observe a bigger correction 
for the $\Omega^-K^+$ mode.
Figure \ref{times} shows the reduced proper time
distribution for each mode after the Monte Carlo correction. For each mode
a likelihood is constructed as the product between the Poisson probability of
observing \( s_{i} \) events in the signal region, when \( n_{i} \) are expected, and the
Poisson probability of observing \( \sum _{i}b_{i} \) background events when
\( B \) are expected. A factor of 2 is included because the sidebands are twice as wide as the signal region.
The likelihood for each category is thus: 
\[ 
L=(\Pi _{i}\frac{n_{i}^{s_{i}}e^{-n_{i}}}{s_{i}!})\times (\frac{(2B)^{\sum _{i}b_{i}}e^{-2B}}{(\sum _{i}b_{i})!}).\]
 To measure \( \tau  \), we maximize the combined likelihood, which is the
product of the likelihoods for the two decay modes: 
\[
L_{\Xi _{c}^{0}}=L_{\Xi \pi }\times L_{\Omega K}.\]
There are three fit parameters: the lifetime \( \tau  \), and for each mode
the background events $B$. Our fit result is \( \tau =118_{-12}^{+14} \)~fs.
When fitted separately, we measure for the two decay modes
\( \tau(\Xi_c^0 \rightarrow \Xi^- \pi^+) =117_{-13}^{+15} \)~fs
and \( \tau(\Xi_c^0\rightarrow\Omega^- K^+) =122_{-25}^{+37} \)~fs.}{\normalsize \par}
\section{{\normalsize Systematic uncertainty determination}\normalsize}

\begin{figure}
{\par\centering \resizebox*{10cm}{10cm}{\includegraphics{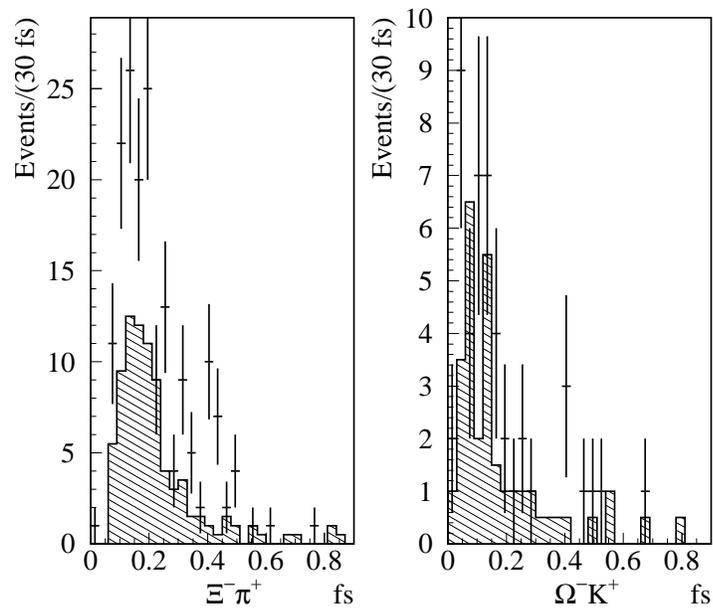}} \par}

\caption{\label{t_conv}Proper time distribution for the two decay modes in the signal region and sideband
region (shaded).}
\end{figure}

{\normalsize The systematic uncertainty was evaluated from a detailed study of different sources:
the Monte Carlo production and detector simulation, the fitting procedure, and the possible intrinsic bias of the method.}{\normalsize \par}
{\normalsize The Monte Carlo correction function may be incorrect if the Monte Carlo poorly simulates the detector or
the production characteristics of the events. For this reason we extract a systematic uncertainty due to the 
Monte Carlo simulation after a careful study of how well the simulation 
matches the data.
We analyze effects from the momentum and transverse momentum of \( \Xi _{c}^{0} \), the flavor (particle and antiparticle)
of the \( \Xi _{c}^{0} \), the decay mode of the \( \Xi _{c}^{0} \), the multiplicity and position of the primary vertex, 
the error on the decay time and length, and the target region silicon strip detector (available for 2/3 of the FOCUS data
set).
To obtain a systematic error, we split the data into statistically
independent samples, on the basis of the production variables and of the mode.
Then we compute the $\chi ^2/{\rm d.o.f.}$ for the hypothesis of consistency of the measurements. If we find 
$\chi ^2/{\rm d.o.f.}>1$, we rescale the errors in order to have 
$\chi ^2/{\rm d.o.f.}=1$, and extract the
systematic error subtracting in quadrature the statistical error from the scaled error of the weighted 
average of the independent measurements.
This method is based on the $S$-factor method of the Particle Data Group~\cite{mesont}.
The only significant effects found are due to primary vertex multiplicity and \( \Xi _{c}^{0} \) decay mode from which a
systematic uncertainty of $\sigma(\rm production)=\pm$~3~fs is obtained. }{\normalsize \par}
{\normalsize The measurement we report is from a particular choice for the
fit parameters and fitting technique (namely, the binned maximum likelihood technique).
Varying these fitting conditions provides measurements which are all} \textit{\normalsize a priori }{\normalsize likely.
We calculate a systematic uncertainty by performing a set of lifetime measurements
with different choices for the fitting parameters and the fitting technique. 
To study systematic effects related to our choice of
fitting parameters we varied the location and width of the sidebands, the proper decay time
fitting region and bin size. We also varied the correction function $f(t')$, by
using a linear fit instead of the bin values.}{\normalsize \par}
{\begin{table}
\begin{center}
\begin{tabular}{c c }
\hline\hline
\rule[-0.2cm]{0pt}{0.1cm}
Contribution     & Systematic (fs)                \\
\hline
\rule{0pt}{0.5cm}
Production     & $\pm$~3   \\
Fit            & $\pm$~4   \\
Method         & $\pm$~2   \\
\hline
Total          & $\pm$~5 \\
\hline
\end{tabular}
\end{center}
\caption{Contributions to the systematic uncertainty.}
\label{tb:sommario}
\end{table}}
{\normalsize Since the proper time resolution (about 40~fs for the \( \Xi ^{-}\pi^{+} \) mode and 80~fs for 
the \( \Omega ^{-}K^{+} \) mode) is close to the \( \Xi ^{0}_c \) lifetime, we decided to perform the measurement
also with a convolved binned likelihood method~\cite{omegac0} to check the fitting technique. In this method, 
the exponential decay is convolved with the smearing due to the time resolution. We used the Monte Carlo
convolution correction function
\( F(t_{j}^ {\rm gen},t_{i}^ {\rm rec}) \), which represents the probability of reconstructing
an event in the decay time bin $i$ when the true decay time bin is $j$.
The likelihood is constructed in the same way as explained before,
but the number of expected events in the \( i^{\rm th} \) bin becomes: 
\[
n_{i}=S\frac{\sum _{j}f(t^{\rm gen}_{j},t^{\rm rec}_{i})e^{-t_{j}/\tau }}{\sum _{i}\sum _{j}f(t^{\rm gen}_{j}
,t^{\rm rec}_{i})
e^{-t_{j}/\tau }}
+B\frac{b_{i}}{\sum _{i}b_{i}}.\]
Instead of the reduced proper decay time, the proper decay time is fit. Fig.~\ref{t_conv}
shows the proper time distributions. 
The lifetime result of the convolved method is
then considered as a further fit variant.
The systematic uncertainty due to the fit variants is given by the variance of the set of 
measurements. We find $\sigma \rm (fit)=\pm$~4~fs.}{\normalsize \par}

{\normalsize The Monte Carlo simulation uses an input lifetime given by the current
world average reported from the PDG~\cite{mesont} of \( \tau (\Xi_c^0)= 98\)~fs. We investigate the range of good
performance of the fitting method when the input lifetime for the correction function is 100~fs
with a mini Monte Carlo study. This consists of simulating a large number of samples with the same statistics as
our data (both for signal and background events). We obtain events from
exponential decaying populations with a given lifetime for the signal and for the background. 
We obtain the background lifetime by 
fitting the data sideband distributions to an exponential.
For the signal events
we study a wide range of lifetime values (from 10~fs to 170~fs). 
Care is taken to account for the time resolution. 
The method works with an accuracy better than 10\% for lifetimes greater than 55~fs. We study the accuracy in
the range of the $\Xi_c^0$ measured lifetime, and measure a systematic
uncertainty \( \sigma \rm (method)\) of $\pm$~2~fs. The mini Monte Carlo study also validates our estimate of the fit statistical
error.}{\normalsize \par}
{\normalsize The total systematic uncertainty is obtained by adding in quadrature the
contributions from the three independent sources.
We thus find \( \sigma \rm (sys)=\pm\)~5~fs. See Table \ref{tb:sommario} for a summary of the contributions and the
total systematic uncertainty. }{\normalsize \par}

\section{{\normalsize Conclusions}\normalsize }

\begin{figure}
{\par\centering \resizebox*{10cm}{10cm}{\includegraphics{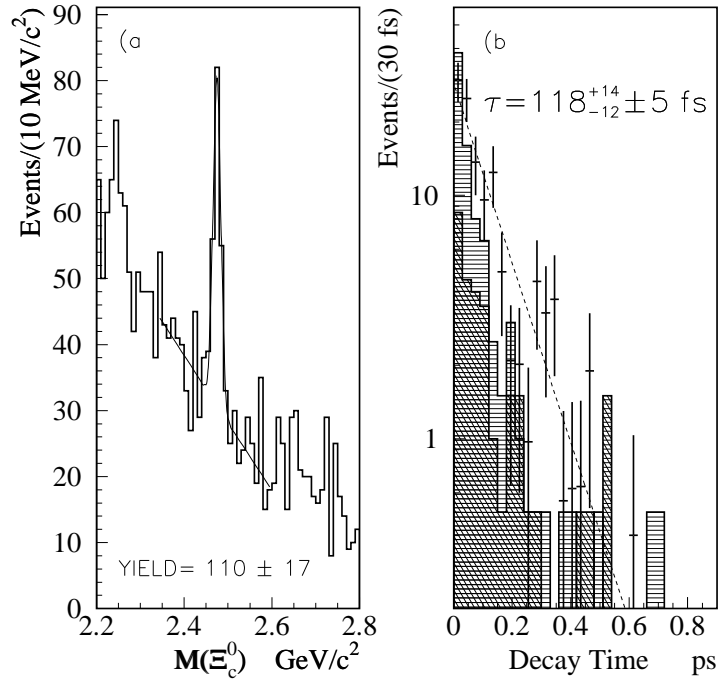}} \par}

\caption{\label{sum_times} Mass and lifetime distributions for the combined sample. a) Invariant
mass distribution; fit with a Gaussian for the signal and a first order polynomial for the background.
b) Monte Carlo corrected and sideband subtracted reduced proper time distribution for the signal region
(points).
The histograms with different shades are the time distributions for the sidebands for the two decay 
modes (lighter shade: \( \Xi ^{-}\pi ^{+} \),
darker shade: \(\Omega ^{-}K^{+} \)). The dashed line shows the result of the fit.}
\end{figure}

{\normalsize We have measured the lifetime of the charmed baryon \( \Xi _{c}^{0} \), fully reconstructing
the decays \( \Xi _{c}^{0}\rightarrow \Xi ^{-}\pi ^{+} \), and \( \Xi _{c}^{0}\rightarrow \Omega ^{-}K^{+} \). 
Figure \ref{sum_times}a shows the invariant mass plot for the combined sample.  
From the \( 110\pm 17 \) 
reconstructed events we measure \( \tau (\Xi _{c}^{0})=118^{+14}_{-12}~\rm (stat)\pm 5~\rm (sys) \)~fs.
Figure \ref{sum_times}b shows the Monte Carlo
corrected sideband subtracted proper time
distribution for the combined sample for the signal region and sideband region.
This measurement greatly improves upon the accuracy of previous measurements, reducing the percentage error 
on the \( \Xi _{c}^{0} \) lifetime from 20\%
to 10\%. }{\normalsize \par}

\section{Acknowledgements}
We wish to acknowledge the assistance of the staffs of Fermi National
Accelerator Laboratory, the INFN of Italy, and the physics departments of the
collaborating institutions. This research was supported in part by the U.~S.
National Science Foundation, the U.~S. Department of Energy, the Italian
Istituto Nazionale di Fisica Nucleare and Ministero dell'Universit\`a e della
Ricerca Scientifica e Tecnologica, the Brazilian Conselho Nacional de
Desenvolvimento Cient\'{\i}fico e Tecnol\'ogico, CONACyT-M\'exico, the Korean
Ministry of Education, and the Korean Science and Engineering Foundation.

\clearpage


\end{document}